%% file: ijcai25.tex
\documentclass{article}
\usepackage{ijcai25}

\usepackage{times}
\usepackage{soul}
\usepackage{url}
\usepackage[hidelinks]{hyperref}
\usepackage[utf8]{inputenc}
\usepackage[small]{caption}
\usepackage{graphicx}
\usepackage{amsmath}
\usepackage{amsthm}
\usepackage{booktabs}
\usepackage{algorithm}
\usepackage{algorithmic}
\usepackage[switch]{lineno}

\usepackage{hyperref}       
\usepackage{url}            
\usepackage{booktabs}       
\usepackage{amsfonts}       
\usepackage{nicefrac}       
\usepackage{microtype}      
\usepackage{xcolor}         
\usepackage{graphicx}
\usepackage{amsmath}
\usepackage{booktabs}
\usepackage{makecell}
\newtheorem{theorem}{Theorem}[section]
\newtheorem{definition}[theorem]{Definition}
\usepackage{threeparttable}
\usepackage{multirow}

\title{QiMeng-CPU-v2: Automated Superscalar Processor Design
by Learning Data Dependencies}

\author{
    Shuyao Cheng$^{1}$, Rui Zhang$^{1}$, Wenkai He$^{1,2,3}$, Pengwei Jin$^{1,2,3}$, Chongxiao Li$^{1,2,3}$, Zidong Du$^{1,4}$, Xing Hu$^{1,4}$, Yifan Hao$^{1}$, Guanglin Xu$^{1}$, Yuanbo Wen$^{1}$, Ling Li$^{5}$, Qi Guo$^{1}$, Yunji Chen$^{1,2}$\footnote{Yunji Chen (cyj@ict.ac.cn) is the corresponding author.}
    \affiliations
   $^1$State Key Lab of Processors, Institute of Computing Technology, Chinese Academy of Sciences \\
$^2$University of Chinese Academy of Sciences \\
$^3$Cambricon Technologies  \\
$^4$Shanghai Innovation Center for Processor Technologies \\
$^5$Institute of Software, Chinese Academy of Sciences \\
}

\begin{document}

\maketitle

\input{Tex/0-abstract}

\input{Tex/1-introduction}
\input{Tex/2-problem_statement}
\input{Tex/3-design_overview}
\input{Tex/4-methodology}

\input{Tex/5-evaluation}
\input{Tex/6-related_work}

\input{Tex/7-conclusion}


\bibliography{ijcai25}

\clearpage

\end{document}

%% file: Tex/0-abstract.tex
\begin{abstract}

Automated processor design, which can significantly reduce human efforts and accelerate design cycles, has received considerable attention. While recent advancements have automatically designed single-cycle processors that execute one instruction per cycle, their performance cannot compete with modern superscalar processors that execute multiple instructions per cycle. Previous methods fail on superscalar processor design because they cannot address inter-instruction data dependencies,  leading to inefficient sequential instruction execution.
 
This paper proposes a novel approach to automatically designing superscalar processors using a hardware-friendly model called the \emph{Stateful Binary Speculation Diagram} (State-BSD). We observe that processor parallelism can be enhanced through on-the-fly inter-instruction dependent data predictors, reusing the processor's internal states to learn the data dependency. To meet the challenge of both hardware-resource limitation and design functional correctness, State-BSD consists of two components: 1) a lightweight state-selector trained by simulated annealing method to detect the most reusable processor states and store them in a small buffer; and 2) a highly precise state-speculator trained by BSD expansion method to predict the inter-instruction dependent data using the selected states. It is the first work to achieve the automated superscalar processor design, i.e. \texttt{QiMeng-CPU-v2}, which improves the performance by about $380\times$ than the state-of-the-art automated design and is comparable to human-designed superscalar processors such as ARM Cortex A53.

\end{abstract}


%% file: Tex/1-introduction.tex
\section{Introduction}

Designing processors automatically has long been a very attractive problem without much success, as processors are considered the most complicated man-made objects. 
Recently, there have been significant advancements in automated processor design where state-of-the-art machine learning methods successfully design single-cycle processors~\cite{cheng2023pushing,fu2023gpt4aigchip,blocklove2023chip}. 
Single-cycle processors, however, are inherently limited in performance as they can only execute one instruction per cycle. 
Therefore, the design of single-cycle processors is merely the initial step toward modern practical processors—precisely, superscalar processors— capable of executing multiple instructions per cycle. 
Currently, existing automated design methods cannot achieve superscalar processor design because they disregard the internal states in the processor.
These internal states contain crucial information about the inter-instruction dependencies, which are inevitable for allowing instruction-level parallelism.

\begin{figure}[t]
  \centering
  \includegraphics[width=\columnwidth]{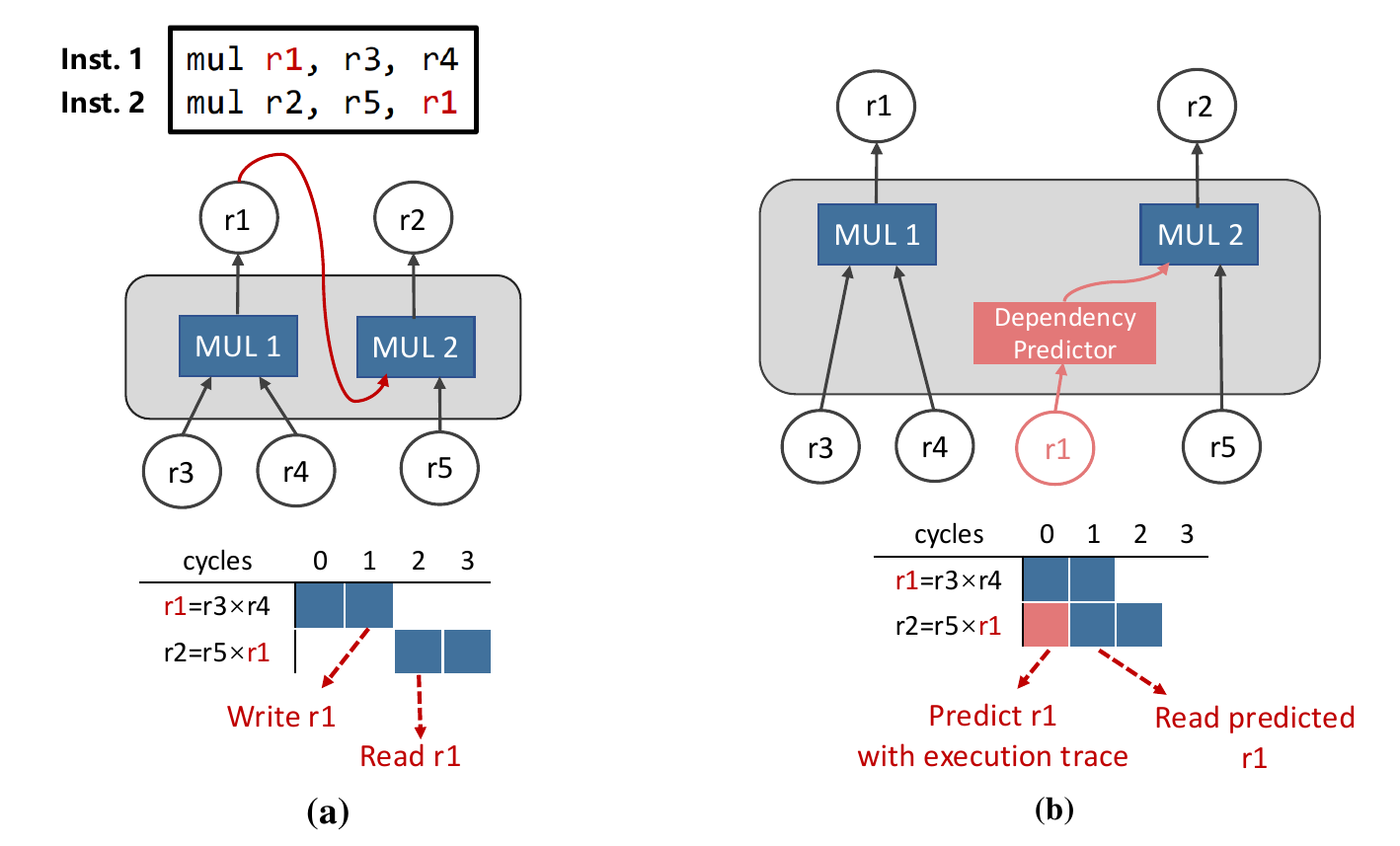}
  \caption{\textbf{An example of breaking data dependencies by a predictor in the superscalar processor design.} 
  \textbf{(a)} For the program slice (Inst.1 and Inst.2) under Read-after-Write (RAW) dependency, without any predictors, these two instructions have to execute one after another.  
\textbf{(b)} With the data dependency predictor, the predicted data of register $r1$ can be obtained in advance by Inst.2 so that it is no longer blocked by Inst.1. 
}
  
  \label{fig:prediction}
  \vspace{-5pt}
\end{figure}

According to the human design paradigm, inter-instruction dependencies can be disentangled by predicting the dependent data on the fly. 
Specifically, a predictor is deployed on the hardware to 1) store the internal processor states in a buffer and 2) reuse them to predict dependent state transitions, thereby transmitting the dependent data in advance. 
After that, instructions with dependency can be executed in parallel for higher performance. 
However, previous predictors designed by human experts require massive manual efforts to solve even a tiny fraction of the data dependencies, inhibiting the automatic design. 
Thus, learning a data dependency predictor is key to achieving automated superscalar processor design.

Learning an on-the-fly dependent data predictor is challenging in two-fold, i.e., lightweight and high predicting precision.
First, the designed predictor must be lightweight to predict in real-time, requiring predictions to be made within just 1-2 clock cycles in nanoseconds, despite the vast and high-dimensional internal state space. 
This complexity limit makes the current time-series prediction methods based on neural networks, such as RNN, LSTM, and Transformer, ineffective~\cite{han2019review}.
Second, the predictor's precision must be high enough to reach 100\%-precise,  as the designed predictor cannot be modified after the processor is manufactured. 
If the predictor predicts dependent data wrong, the design functionality is incorrect and cannot pass the strict verification process for processor design.


In this paper, we propose to automatically design a superscalar processor by learning data dependencies with a novel hardware-friendly machine learning model \emph{Stateful Binary Speculation Diagram} (State-BSD). Surpassing the state-of-the-art automated processor design method Binary Speculation Diagram (BSD), State-BSD captures the inter-instruction dependencies with a small set of reusable states. In this way, it can be trained to design a highly precise on-the-fly predictor in the processor for instruction-level parallelism, consisting of 1) a lightweight state-selector to select the reusable processor's internal states and store them in a small buffer and 2) a highly precise state-speculator to accurately speculate the state transition and predict the dependent data. More specifically, to meet the lightweight challenge, a small set of states are selected from the vast high-dimensional processor internal-state space by simulated annealing, and stored in a small buffer for further reuse. To meet the precision challenge, the state-speculator uses the Binary Speculation Diagram method to precisely speculate the predictable dependencies with buffered states. According to the human design paradigm, the dependent data can be separated into predictable dependencies and unpredictable dependencies. The state-speculator can 100\%-precise predict the predictable dependencies with BSD expansion, which is proved in the previous work~\cite{cheng2023pushing}, meanwhile bypassing the unpredictable dependencies to avoid the potential functional errors caused by misprediction. With State-BSD, we can automatically design an on-the-fly predictor for instruction-level parallelism, in which the predictable dependent instructions can be paralleled by predicting the dependent data, as shown in Figure~\ref{fig:prediction}.

The proposed automated processor design framework is applied on a large-scale real-world RISCV-32IA CPU, \texttt{QiMeng-CPU-v2}. The performance of the designed CPU significantly outperforms the state-of-the-art automated design CPUs~\cite{cheng2023pushing} by 382$\times$ and is comparable to the modern human-design superscalar processor, i.e., ARM Cortex A53. It has already been validated on the FPGA to successfully run real-world programs including the Linux Operating System and SPEC Benchmarks, and we are taping it out with 28nm technology.
To our best knowledge, this is the world’s first superscalar CPU automatically designed by AI.

This paper makes the following contributions:

\begin{itemize}
\item  We formulate the automated superscalar processor design as a machine-learning problem for predicting the dependent data for instruction-level parallel.
\item 
we propose a novel model State-BSD to capture the inter-instruction dependencies with processor states to learn the data dependencies. 
\item We automatically design a superscalar RISCV-32IA CPU,  i.e. \texttt{QiMeng-CPU-v2}, with State-BSD. The designed processor outperforms the state-of-the-art automated design methods by $382\times$ and is comparable to modern human-designed CPUs.
\end{itemize}

%% file: Tex/2-problem_statement.tex
\section{Problem Statement}


We first formalize the definition of the superscalar processor design and demonstrate how inter-instruction dependencies decrease the processor performance. Then, we transfer the superscalar processor design problem to a machine learning problem, i.e., learning to predict inter-instruction dependent data with processor states. 
We demonstrate that a high-precision predictor can ensure the functional correctness of the superscalar processor design and optimize its performance.

\subsection{Superscalar Processor Design}
 
A processor can be regarded as a finite-state machine that functionally transitions from a current state to the next state, triggered by the input instructions.
So, we define the functional specification of a single-cycle processor as follows:
\begin{definition} \textbf{(Single-cycle Processor)}
    Given an state space set $\Psi=\{\{0,1\}^k\}$, $k$ is the dimension of the state space; an input instruction space set $\Phi=\{\{0,1\}^m\}$, $m$ is the dimension of the input instruction space. The functional specification of a single-cycle processor is
    \begin{equation}
            \mathrm{Processor}:\Psi \times \Phi \rightarrow \Psi.
    \end{equation}
\end{definition}
Correspondingly, a $\mathrm{Program}$ contains a sequence of input instructions $P=[\phi^1,\phi^2,\cdots,\phi^n],\phi^i \in \Phi, i=1,2,\cdots,n$, and an initialized processor state $\psi^0$. The specification defines the processor's functionality, executing the program and transitions as a trace $\psi^0,\psi^1,\cdots,\psi^{n}$, $\psi^i\in\Psi, i=0,1,\cdots,n$.

According to these definitions, we can evaluate the performance of processor design, i.e., the average execution latency of arbitrary instructions, with cycles per instruction (CPI). 
As a single-cycle processor cannot parallel the transition, we define $L(\psi^i,\phi^i)$ as the latency of the state $\psi^i$ with an instruction $\phi^i$, and thus the average execution latency of the single-cycle processor for the program is $\mathrm{CPI} (\mathrm{Processor}) = \sum_{i=0}^n L(\psi^i,\phi^i)/n$. 

Since single-cycle processors can only execute one instruction at the same time, their performance is limited. On the other hand, superscalar processors can execute multiple input instructions in one single clock cycle and conceal the execution latency, defined as:
\begin{definition} \textbf{(Superscalar Processor)}
Given an state space set $\Psi=\{\{0,1\}^k\}$, $k$ is the dimension of the state space; an input instruction space set $\Phi=\{\{0,1\}^m\}$, $m$ is the dimension of the input instruction space; a maximum parallelism degree $p$. The functional specification of a superscalar Processor is
\begin{equation}
{\mathrm{Processor}_s}:\Psi \times {\Phi}^p \rightarrow \Psi.
\end{equation}
\end{definition}
In this way, the latency of the superscalar processor with state $\psi^i$ and the next state $\psi^{i+1}$ can be defined as $L_s(\psi^i,\phi^i)$, which is not greater than the single-cycle processor $L(\psi^i,\phi^i)$. 
In the superscalar processor, $L_s(\psi^i,\phi^i) = L(\psi^i,\phi^i)$ only if the instruction cannot be paralleled, i.e., cannot execute until the last instruction finishes executing, such as data is occupied due to data dependency, or all the execution units are occupied.
So the average execution latency of the superscalar processor is $\mathrm{CPI}~ (\mathrm{Processor_s}) = \sum_{i=0}^n L_s(\psi^i,\phi^i)/n \leq \mathrm{CPI}~(\mathrm{Processor})$.
Precisely, the latency $L_s$ is determined by data dependencies, especially the Read-after-Write ones. According to this definition, $L_s \geq L / p$ reaches the minimum if and only if there are no dependencies in the execution trace. With more dependent data handled by the predictor, the performance of the superscalar processor improves. In the next section, we further quantify how the dependent data predictor affects the superscalar processor performance. 

\subsection{Managing Inter-Instruction Dependencies:   Learning a High-precision Predictor}

Inter-instruction data dependencies can be resolved by predicting the dependent data with a predictor. By transmitting the dependent data in advance, dependent instructions can be executed in parallel, thereby reducing execution latency. In this section, we analyze the performance improvement gained by the data dependency predictor.

Inspired by the human design paradigm, we categorize all dependent data into two groups: predictable data and unpredictable data. Some dependencies may be extremely challenging to predict, so the human-designed predictors strike a balance between the design cost and performance, intentionally bypassing dependencies deemed ``unpredictable'' to avoid the risks associated with potential mispredictions. As a result, the predictor is measured by two primary metrics: coverage ($C$) and precision ($Pr$). 
We demonstrate that design correctness is affected by the precision $Pr$, while the design performance of the superscalar processor is affected by $C$.

\paragraph{The design correctness of the superscalar processor.} 
We first define the positive/negative ($P/N$) and true/false ($T/F$) in the dependent data predictor and then define the predictor's precision and explain how precision impacts design correctness. For dependent data, positive/negative ($P/N$)  indicates whether the predictor outputs a prediction for the processor ($P$) or considers it as unpredictable($N$). True/false ($T/F$)  indicates whether the output of the predictor is the same as the perfect-predicting oracle, i.e., the prediction is $T$ satisfying the following two conditions, and $F$ otherwise: 1) if the data is predictable, the predictor outputs the same data as the oracle, and 2) if the data is unpredictable, the predictor does not output a predicting data.
The precision of the predictor $Pr$ is the True Positive prediction in all the positive predictions (including the True Positive ($TP$) and False Positive ($FP$) predictions), i.e., $Pr = TP/(TP+FP)$. 
﻿
Precision is the principal target in training the predictor, mainly due to the imbalance between the two types of mispredictions: the $FP$ and $FN$, where $FP$ costs much more than $FN$ and is not affordable in the superscalar processor design. A False Positive ($FP$) occurs when the predicting result differs from the perfect-predicting oracle, meaning it incorrectly identifies unpredictable, dependent data as predictable or transmits the wrong prediction to the processor. In this way, the processor can be functionally wrong and cannot pass the functional verification. On the other hand,  a False Negative ($FN$) occurs if the dependent data is predictable but the predictor mis-consider it as unpredictable. In this way, the processor stalls the execution redundantly with one clock period performance loss while keeping the design correct. To design an accurate superscalar processor, the precision of the dependent predictor needs to be 100\%; Given the disproportionate impact of false positives ($FP$) compared to false negatives ($FN$),  the proposed method trades off the predictors recall $Re=TP/(TP+FN)$ for a 100\% precision $Pr = TP/(TP+FP)$, guaranteeing the design correctness of the superscalar processor.

\paragraph{The design performance of the superscalar processor.}
The design performance of the superscalar processor is affected by the coverage $C$ of the predictor, i.e. the number of the predictable data output from the predictor of all the possible dependent data. 
With a data dependency in the processor state $\psi^i$ with an instruction $\phi^i$, the latency $L_s(\psi^i,\phi^i) = L(\psi^i,\phi^i)$ if there is no prediction in the superscalar processor.  With the lightweight predictor successfully predicting the dependent data, the serial execution can be paralleled, and reduce the execution latency from $L(\psi^i,\phi^i)$ to the prediction latency $L_p$, $L_p < L(\psi^i,\phi^i)$. So the more dependent data is correctly predicted, the more performance improvement in the superscalar processor. 
In this way, the coverage of the prediction becomes the key to the automated superscalar processor design performance. For a superscalar processor with predictor $\mathrm{Processor}_p$, if the coverage of the predictor is $C$, then the average performance  $\mathrm{CPI}~(\mathrm{Processor}_p)$ is optimized,
$
          \mathrm{CPI}(\mathrm{Processor}_p)  = L_p \times \mathrm{C} 
         + \mathrm{CPI}(\mathrm{Processor})\times(1-\mathrm{C})) 
            \leq \mathrm{CPI}(\mathrm{Processor})
$

%% file: Tex/3-design_overview.tex
\begin{figure*}[t]
  \centering
  \includegraphics[width=\textwidth]{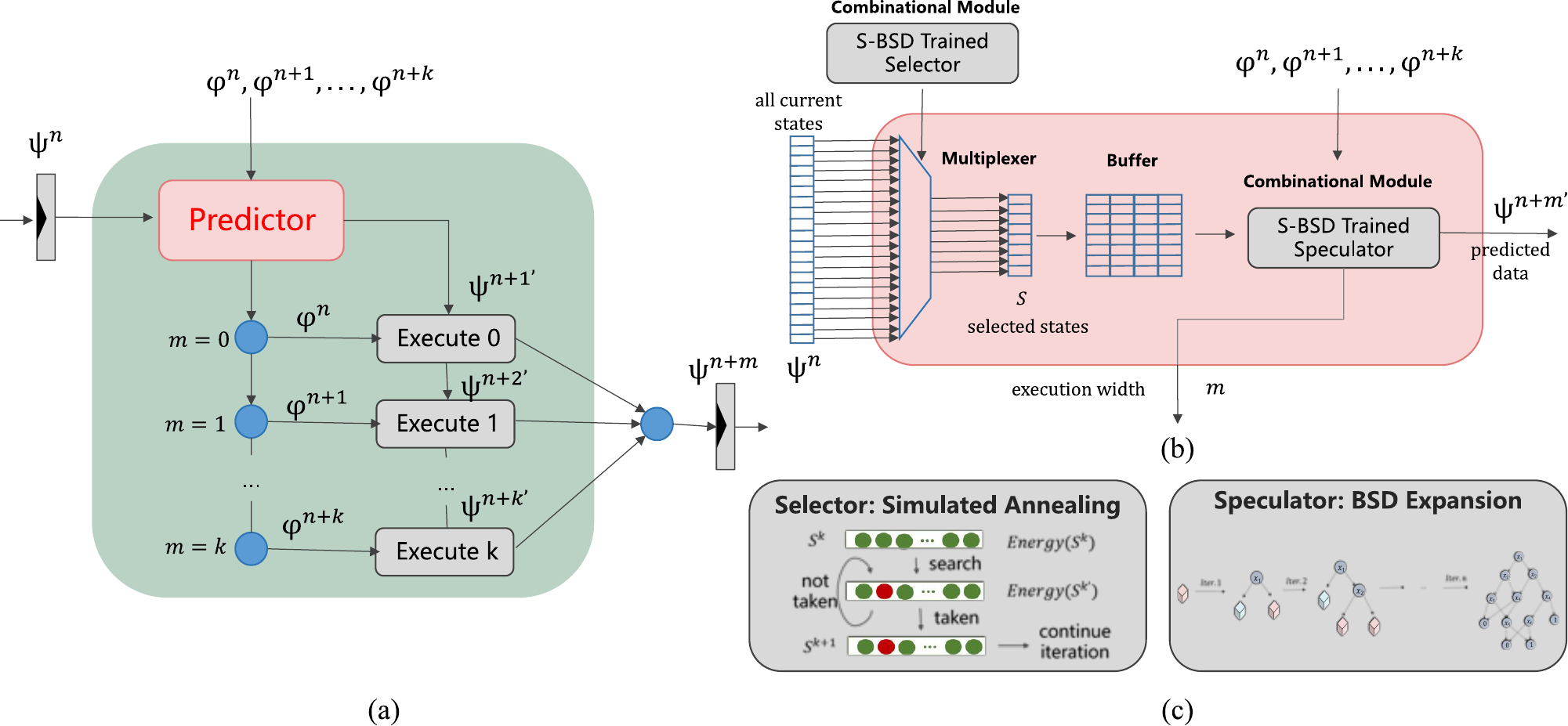}
   \vspace{-10pt}
  \caption{\textbf{The design overview.} (a) The automated designed superscalar processor. It consists of a predictor automatically designed by State-BSD and $k$ execution units. Every clock cycle, the predictor outputs $m$ indicating how many instructions can be execute in parallel, and $\Psi$ indicating the predicted dependent data. (b) The hardware implement of the dependency predictor, consisting of two components: the state-selector and state-speculator. The state-selector is a multiplexer with small buffer, and the state-speculator is a high-precision combinational module. (c) Train the dependency predictor with State-BSD: The selector is trained by simulated annealing, and the speculator is trained by BSD expansion.
  }
  \label{fig:State-BSD}
 \end{figure*}

\section{Methodology}

Based on the problem statement, we propose an automated superscalar processor design flow with a high-precision dependent data predictor, which is learned with a novel hardware-friendly model \textit{Stateful Binary Speculation Diagram (State-BSD)}. In this section, we 1) first illustrate the design overview showing how to design the superscalar processor with dependent data predictor, and 2) then introduce the details about how to train the predictor and implement it in the hardware.

\subsection{Design Overview}

In this section, we provide an overview of the design, specifically focusing on how the superscalar processor is constructed with a learned data predictor.
 
 The input of the design flow is a single-cycle processor, which can be automatically designed by the previous works~\cite{cheng2023pushing}, and the output of the design flow is a superscalar processor. The design flow designs a dependency predictor to control the instruction-level parallelism with multiple execution units, as illustrated in Fig.~\ref{fig:State-BSD} (a). For a processor with $n$ execution units, in the clock cycle, the predictor checks the following $n$ instructions to be executed and whether their operands can be predicted. If the predictor can predict all the data of the following $m$ instructions ($m \leq n$), the superscalar processor executes $m$ instructions in parallel.

Following the paradigm of human-designed processors, we identify three types of dependency potentials:  the data dependency in the load/store unit for memory (MEM), the data dependency in the program counter for instructions (PC), and the data dependency in general purpose registers (GPR). Accordingly, we propose three sub-predictors, each dedicated to one of these data dependencies.
For each sub-predictor, its input is the instruction $\phi$, the current processor states $\psi$, and its output is a control signal $m$ indicating the dependent data in the following $m$ instruction can be predicted by the predictor, along with the predicting data itself. For an ongoing instruction, it can be predicted by the dependent predictor if and only if all three sub-predictor can predict the corresponding potential dependent data.
 
For hardware implementation, we train these three sub-predictors independently (details of the training process are in Section 3.2) and implement them using Verilog code. All the sub-predictors are implemented as a combinational module and then integrated into the processor. The training phase must ensure the design's correctness, as the implementation cannot be modified after the processor is manufactured.

%% file: Tex/4-methodology.tex
\subsection{Train the Data Predictor with State-BSD}

We propose a novel model \emph{Stateful Binary Speculation Diagram} (State-BSD) to train the on-the-fly predictor, which is conducted by a lightweight state-selector and a high-precision state-speculator to predict the dependent data. 
This method ensures the predictor is executed on the fly without too much extra hardware cost while maintaining the processor's functional correctness. 
As shown in Fig~\ref{fig:State-BSD} (b), the state-selector consists of a combinational module that chooses the most reusable states and a small buffer storing the selected states. The state-speculator is a combinational logic module that reuses the buffered states to predict the dependent data. These two components have little hardware overhead compared with the size of the entire processor design. 
 
We propose two training methods for these two components of the State-BSD, as shown in Fig~\ref{fig:State-BSD} (c). The state-selector is trained with a simulated annealing method, searching for the most reusable states to ensure that the predictor maintains effectiveness with only a small state-buffer. The state-speculator is trained with a Binary Speculation Diagram (BSD), which is proved in the previous work that the precision can be boosted to over 100\%, thereby ensuring the design's overall correctness. Although the training method is too complex to implement in the hardware, it runs before the processor is manufactured: only the trained components, the state-selector and state-speculator, are implemented on the hardware.

\textbf{a. State-selector: select the most-reusable states with Simulated Annealing, and store them in a small buffer.} 
The state-selector automatically chooses the most reusable states $S$ for prediction and stores it in the state-buffer. Inspired by the human-designed paradigm, i.e., the \textit{Least Recently Used} (LRU) based methods in the buffer designs, State-BSD aims to select a set of processor states that are reused the most frequently. Furthermore, the reusability of a set of selected states can be measured with a software simulator: given a sequence of instructions, record the dependent data of each instruction and compare it with the selected states; the reusability of the chosen state set is determined by the ratio of dependent data that matches the selected state set to the total dependent data.
 
The state-selector is trained by a simulated annealing method, starting with a randomly selected set of states $S_0$, and iteratively updates the selected state set to optimize the energy function ( i.e. the reusability of the selected states). Finally, the iteration ends when the updating method cannot find another set of the selected states with lower energy. 
In the $k$-th iteration of the simulated annealing, the proposed method aims to update the selected set of states from $S_k$ to $S_{k+1}$ to decrease the energy function $E$. It first randomly samples a set of states $S_{k'}$, which contains a small change on the starting set of states $S_k$, then calculate the energy function $E(S_{k'})$. If $E(S_{k'}) < E(S_{k})$, showing that $S_{k'}$ is a more reusable set of states, $S_{k'}$ is taken to update the selected set of states, i.e. $S_{k+1} = S_{k'}$. Otherwise if $E(S_{k'}) \geq E(S_{k})$, $S_{k'}$ is taken to update the selected set of states with an adaptive probability $P = \mathrm{e}^{-\frac{E(S_{k'}) - E(S_{k})}{T}}$, where $T$ is an adaptive hyperparameter decreases when $k$ increases, showing that $S_{k'}$ is more likely to be taken to update if it has better reusability. If $S_{k'}$ is not taken to update, the proposed method re-samples a new set of selected states with the same sampling method and continues the iteration. If the re-sample time reaches a threshold and $S_k$ eventually cannot be updated, the simulated annealing method finishes and outputs $S_k$ as the final result, i.e., the selected set of most reusable states.
 We then automatically implement the state-selector in the processor hardware design (only the trained selector itself is implemented on hardware; the simulated annealing method is for training and not hardware implemented). The selecting logic is a combinational multiplexer whose input is all the possible internal processor states, and the output is a small set of selected states. The selected state set is stored in a small buffer, and these selected states are used as the input of the state speculator.

\textbf{b. State-speculator: a highly precise combinational module designed with speculative graph expansion.} With the buffered data, the state-speculator aims to predict the dependent data precisely with the reusable data. The state-speculator is a combinational module whose input includes the current input instruction $\phi$ and the selected states in the buffer $S$, and output includes the predicted dependent data $D$ and a single-bit controlling signal $C$ that whether the predicting data is valid. 
During training, since the perfect predicting result can be obtained by simulation, we can obtain a perfect predicting oracle $F: \{\Psi, S\} \rightarrow D$ with only input-output observations. Unfortunately, the perfect predicting oracle is a simulated result and cannot be implemented on the hardware. In this way, our goal is to design a highly precise hardware-friendly speculator $G: \{\Psi, S\} \rightarrow \{D, C\}$ from the input-output observations of the perfect predicting oracle $F$. 
We formalize the problem as follows:
\begin{definition} \textbf{(Highly-Precise Logic Regression)}
     Given a logic function $F:\{\Psi,S\} \rightarrow D$ with only input-output examples, design a logic function $G = \{G_d,G_c\}$, where $G_d: \{\Psi,S\} \rightarrow D$ and $G_c: \{\Psi,S\} \rightarrow \{0,1\}$, such that for every $\psi \in \Psi$, $s \in S$,$(G_d(\psi,s) = F(\psi,s)) \vee (G_c(\psi,s) = 1)$.
\end{definition}

This problem is solved with an existing model called \textit{Binary Speculation Diagram} (BSD)~\cite{cheng2023pushing}. Starting from a root-speculated node, the speculator is trained by iteratively expanding one speculated node into two sub-nodes separated by the selected states in the buffer $S$, according to the Boolean Expansion Theorem~\cite{boole1854}. It is proved that with more nodes expanded, the precision of the speculator increases and can finally reach 100\%~\cite{cheng2023pushing}. The speculator ends training until the precision is 100\%.

After training the state-speculator, the precision of the predictor can be verified with an SMT solver to guarantee the design's correctness. If the design is not formally precise, the state-speculator requires an incremental training phase with more input-output examples until it passes the formal verification. The SMT solver is only used for verification and is not implemented in the processor hardware design.





\subsection{The difference between the BSD and State-BSD.}


The state-of-the-art automated processor design method BSD can only design combinational functions with input-output examples and thus cannot manage the Read-after-Write dependencies in the design. In this way, it can only be used to design single-cycle processors without any invisible states.

State-BSD aims to learn the inter-instruction dependency, and requires the internal processor states during the execution. To fill the gap between the vast state space and limited hardware resources, it selects the most reusable states and stores them in a small buffer. With the buffered data, State-BSD takes advantage of the high-accuracy guarantee from the BSD and designs a high-precision state-speculator. With the extra usage of the state information, State-BSD is an extension of the current BSD method, which can handle the data dependency while maintaining high design accuracy.

%% file: Tex/5-evaluation.tex
\section{Evaluation}

In this section, we evaluate the proposed State-BSD method in terms of the processor design performance, the predictor effectiveness, and the efficiency of the State-BSD components. First, we compare State-BSD with the state-of-the-art automated processor design methods. Second, we compare State-BSD with the human design paradigm. Third, we make a detailed algorithm analysis for the State-BSD, demonstrating the effectiveness of every State-BSD component. 

We evaluate the State-BSD on a large-scale, real-world RISCV-32IA CPU, outperforming the largest-scale processor that state-of-the-art automated methods can design. It is the second version of the automated CPU design after the one proposed in \cite{cheng2023pushing}, and it is called \texttt{QiMeng-CPU-v2}.
Our design is a 4-ALU superscalar design with a 2KB buffer in each predictor.  The CPU is functionally correct on over $10^{12}$ instructions on real-world programs, including Linux System, SPEC CPU Benchmark, and others. 
Additionally, the CPU is taping out with 28nm technology; the hardware characteristics are shown in Table~\ref{tab:hwchar}. 

The designed CPU is evaluated on the standard CPU benchmarks, Dhrystone~\cite{weicker1984dhrystone} and Coremark~\cite{CoreMark}, measured by both the benchmark result and the corresponding Cycles per Instruction (CPI) on both Xilinx Zynq UltraScale+ ZCU104 FPGA and commercial simulation tools. The design and evaluation run on CentOS 7.8 with 2 Intel Xeon Gold 6230 CPUs and 512G Memory. The performance on the FPGA is about CPI=2.3, and the simulated performance is about 6.2M Dhrystones/s. To the best of our knowledge, this paper is the first to achieve an automated superscalar processor design, and it is currently the best-performance processor design by AI.

\begin{table}[t]
  \centering
  \small
  \caption{Hardware characteristics of superscalar processor.}
  \resizebox{\linewidth}{!}{
    \begin{tabular}{lllll}
    \toprule
    \textbf{Component} & \textbf{Area ($um^2$)} & \textbf{(\%)} & \textbf{Power (mW)} & \textbf{(\%)} \\
    \midrule

    Superscalar CPU & 798525.92&100.00&271.84 &100.00\\

    \midrule

    Combinational &715,757.28&89.63&55.28&20.34\\

    Register &82,768.64&10.37&216.56&79.66\\

    \bottomrule   
 \end{tabular}
 }
 \label{tab:hwchar}
\end{table}

\subsection{Comparison with the State-of-the-art Automated Design Methods}

We compare the proposed State-BSD method with the state-of-the-art automated processor design methods, including the Reinforcement Learning~\cite{roy2021prefixrl}, Decision Trees~\cite{chen2020circuit}, LLMs~\cite{blocklove2023chip} and BSD expansion methods~\cite{cheng2023pushing}. The results are shown in Table~\ref{tab:comp}, in which the scale is measured by the number of logic gates, and performance is measured by the Dhrystone score.

\textbf{The CPU design performance.} The CPU design performance is evaluated with its Dhrystone benchmark score, as shown in Table~\ref{tab:comp}. The most state-of-the-art automated designed CPU, \textit{Enlightenment-1}, designed by BSD expansion methods~\cite{cheng2023pushing} is a single-cycle RISCV CPU whose throughput is about 16K Dhrystone/s,  only comparable to an Intel 486 CPU. For other methods, such as Neural Networks and Reinforcement Learning methods, they cannot design accurate enough processors, i.e., accuracy $>99.99999999999\%$~\cite{Bentley01DAC} as the human-design CPUs, so that the throughput is not available testing on standard benchmarks. With our proposed method with State-BSD, the average throughput is about 6.3M Dhrystone/s, about $382\times$ outperforming the state-of-the-art design. Experimental results show that with the ability to use the stateful information in the processor, State-BSD benefits from the dependant data prediction, and outperforms the state-of-the-art methods.

\textbf{The predictor effectiveness.} We evaluate the predicting coverage and precision with state-of-the-art methods. The overall data predicting coverage of the proposed method is 70.47\%, meaning that over 70\% of the instructions can be paralleled with its predecessor, and in these covered data, the precision is 100\%. More details and evaluations about the predictor effectiveness are in section 4.3.
We compare it with the predictor designed by the state-of-the-art methods. However, we notice that all the state-of-the-art methods cannot reach 100\% precision so they cannot design functional-correct superscalar processors.

\textbf{The designing scale.} Table~\ref{tab:comp} shows that the scale of the proposed automatically designed superscalar processor is much larger than those circuits designed by the state-of-the-art methods. The scale limit makes these methods unable to accomplish high-quality designs with more complex structures. By exploring the larger space with complex dependency, State-BSD is capable of not only designing larger-scale processors but also using this ability to improve the design performance.


\begin{figure}[t]
  \centering
  \includegraphics[width=0.8\linewidth]{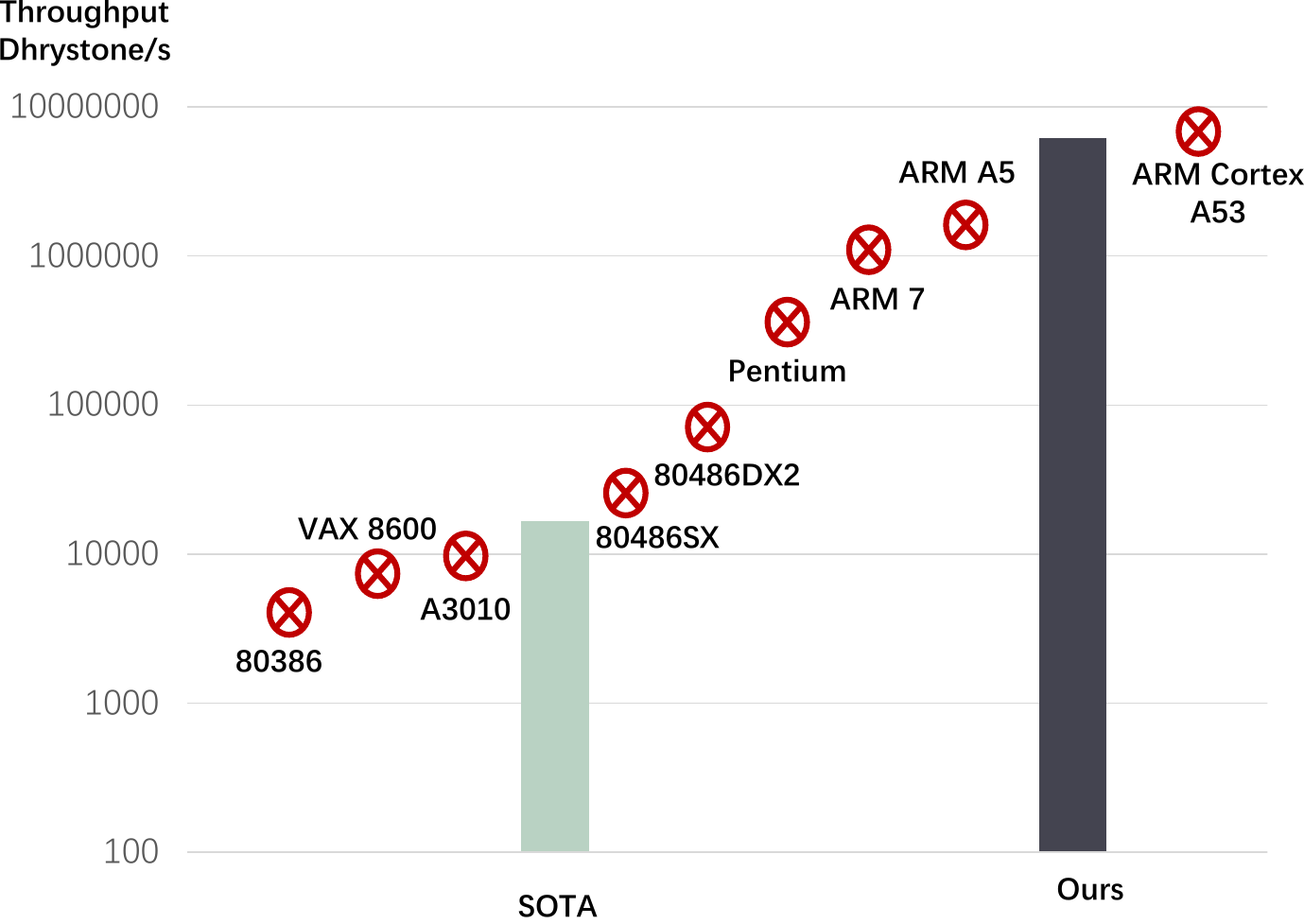}
  \caption{\textbf{CPU performance comparison with the state-of-the-art.} We compare the performance of the automated designed CPU designed by our method and the state-of-the-art. The red data points are human-designed commercial CPUs with similar performance. The result shows that our design is comparable to Arm Cortex A53 (2010s CPU), while the state-of-the-art automated design is only comparable to Intel 486 (1990s CPU).}
  \label{fig:CPU performance}
  \vspace{-10pt}
\end{figure}

\begin{table}[t]
\centering
\small
\caption{Comparison with the state-of-the-art.}
\resizebox{\linewidth}{!}{
\begin{tabular}{cccc}
     \toprule
                Methods    &  \textbf{Target Circuit}    &\textbf{Scale}   &\textbf{Performance}    \\
     \hline
RL & Adder~\cite{roy2021prefixrl} & 118 & NA  \\ \hline
DT & Circuit Modules~\cite{chen2020circuit} & 186 & NA   \\ \hline
EL & Circuit Modules~\cite{rai2021logic} & $\sim$2500 & NA  \\ \hline
LLM & 8-bit CPU~\cite{blocklove2023chip} & 999 & NA  \\ \hline
BSD & RISCV-32 CPU~\cite{cheng2023pushing} & $\sim$4 Million & $1.62\times 10^4$  \\ \hline
\textbf{State-BSD (Ours)} & Superscalar CPU & $\sim17$ Million & $6.29\times10^6$ \\

     \bottomrule
\end{tabular}
}
\vspace{-0pt}
\label{tab:comp}
\end{table}

\subsection{Comparison with the human design}
 
We compare the design performance of State-BSD with both the overall human-design CPU and the human-design predictor within the same computer architecture. The experimental results show that State-BSD significantly narrows the gap between automated and human design abilities.
 
\textbf{The CPU design performance.} We compare the performance of the automated processor design along the commercial CPU developing timeline with the official ARM CPU benchmark Dhrystone. 
Experimental results in Figure~\ref{fig:CPU performance} show that our method significantly improves the automated design ability from the CPUs in the 1990s (16K Dhrystone/s) to the CPUs in the 2010s (6.2M Dhrystone/s).
 
\textbf{The predictor effectiveness.} For a human-designed superscalar processor with high-quality predictors of the same architecture, the average coverage of the predictor is 74.9\%. 
The results show that State-BSD exploits most of the optimization potentials, i.e., 94\%  possible parallel instructions human designers find while eliminating massive human design effort. 


\begin{table}[t]
\centering
\footnotesize
\caption{The effectiveness of the state-selector.}
\begin{tabular}{ccccccc}
     \toprule
                   &   Performance  & &Coverage& Precision \\
     \hline
    w/o   & $4.400\times 10^4$ &  & $0$ &N/A\\
    128B  & $9.539\times 10^4$ & $2.17\times$ & $12.8\% $ & $100\%$\\
    256B  & $1.121\times 10^5$ & $2.55\times$ & $16.5\% $ & $100\%$\\
    512B  & $2.425\times 10^5$ & $5.51\times$ & $29.7\% $ & $100\%$\\
    1KB   & $3.368\times 10^5$ & $7.66\times$ & $35.7\%$ & $100\%$\\
    2KB   & $6.285\times 10^5$ & $142\times$ & $86.7\% $ & $100\%$\\
    4KB   & $6.289\times 10^5$ & $143\times$ & $86.8\% $ & $100\%$\\
    Perfect  & $1.339\times 10^6$ & $304\times$ & N/A & N/A\\

     \bottomrule
\end{tabular}
\vspace{-0pt}
\label{tab:evaluation}
\end{table}
\subsection{State-BSD Algorithm Effectiveness Analysis}

In this section, we evaluate the efficiency of both components in the State-BSD, i.e., the state-selector and state-speculator. We compare the performance of the different components with the intuitive design for ablation study, and with the best-in-theory perfect predictor to analyze its efficiency. The best-in-theory perfect predictor is achieved by simulating the exact execution, so that it can only used for simulation and cannot be implemented on the hardware.




\textbf{State-selector.} We compare the proposed state-selector with/ without the state-selector for ablation study, and evaluate its effectiveness with the best-in-theory perfect predictor.
Without the selector, i.e. predicting with the current instruction without any stored states, the design performance of the superscalar processor is 44.0K Dhrystone/s, and it cannot predict any data dependency. Table~\ref{tab:evaluation} shows the effectiveness of the state-selector, that the design performance increases with larger buffers, showing that the selector can increasingly find reusable processor states. Compared with the best-in-theory perfect predictor, the proposed method can find 86.7\% of the overall dependent data with only 2KB buffers. Besides, the buffer size cannot improve the design performance infinitely.

\textbf{State-speculator.} We compare the proposed design with/ without speculator, and the best-in-theory perfect predictor to analyze its effectiveness. Without a highly-precise speculator, the predictor need to be conservative that it does not predict any dependent data to avoid functional errors. Table~\ref{tab:ablation} shows the effectiveness of the state-speculator in each of the three sub-predictor for PC/GPR/MEM.
The result shows that each of the state-speculator is effective to improve the performance close to the best-in-theory predictor, with 100\% precision.


\begin{table}[t]
\centering
\small
\caption{The effectiveness of the state-speculator.}
 \vspace{-5pt}
\resizebox{1.0\linewidth}{!}{ \hspace{-0.2cm}
\begin{tabular}{cccccccc}
     \toprule
                &\multicolumn{3}{c}{Performance}  &\multirow{2}{*}{Precision}\\
                    &  without & with & perfect    \\
                   
     \hline
    PC   & $9.655\times 10^6$ & $9.962\times 10^6$ & $1.016\times 10^7$  &100\%\\
    MEM  & $1.622\times 10^4$ & $3.986\times 10^6$ & $3.986\times 10^6$  &100\%\\
    GPR  & $5.973\times 10^6$ & $6.267\times 10^6$ & $6.647\times 10^6$  &100\%\\

     \bottomrule
\end{tabular}
}
\vspace{-10pt}
\label{tab:ablation}
\end{table}

%% file: Tex/6-related_work.tex
\section{Related Work}

In 1950, Claude Shannon envisioned computers to ``design electrical filters and relay circuits"~\cite{shannon1988chess}.
Since then, various automated circuit design methods have been developed over the decades, while these methods cannot yet design high-performance processors. In this section, we introduce both the history of the automated processor design and the corresponding human-design paradigm.



\subsection{Automated processor design}

With the advent of machine learning techniques, it has become feasible to design functional single-cycle processors despite the performance inefficiency.
The state-of-the-art methods for automated processor design can design large-scale CPUs or domain-specific accelerators. For example, Blocklove et al. proposed to generate an 8-bit CPU with intensive interactions between the large language model (LLM) and human engineers, the design is quite small with only $\sim1000$ logic gates and without any parallel structures~\cite{blocklove2023chip}. There are also methods designing Verilog code for similar-sized circuit but not the processor with LLMs~\cite{liu2024rtlcoder,vijayaraghavan2024circuitsynth,xiao2024prefixllm} and NNs~\cite{zhang2019circuit}.
Fu et al.~\cite{fu2023gpt4aigchip} propose a new design automation pipeline for the AI accelerator with LLMs. Cheng et al.~\cite{cheng2023pushing} proposed to generate an industrial-scale CPU and layout the design, while the performance is only similar to the Intel 486 CPU. However, the state-of-the-art methods automatically design the single-cycle processor functionally, but cannot design the parallel architecture for comparable performance optimization with the modern CPU.

\subsection{Superscalar CPU design with Value Prediction}

Predicting the dependent data, i.e. value prediction,  is a classic manual design method to apply instruction-level parallelism in the processor design.
In 1996, it was first proposed in the processor design~\cite{lipasti1996value,gabbay1996speculative}. In 1998, it was applied separately in two different domains, the load-store unit(MEM), and the general processing registers(GPR), on a simple RISCV CPU~\cite{marcuello1999value}. Value prediction has been studied for decades, and it becomes more important when the manufacturing technological progress slows down~\cite{hennessy2019new}. After 2014, the value prediction method became a practical CPU design approach and is applied in a modern CPU with complex ILP optimizations such as Out-of-order, and can be integrated into the state-of-the-art human-designed CPU~\cite{perais2014practical}. More accurate value predictors are proposed to increase the prediction coverage, for better instruction-level parallelism in the superscalar processor design~\cite{orosa2018avpp,bandishte2020focused,perais2021leveraging}.
However, all these value prediction methods require massive human effort to design the superscalar processor, as current automated design methods cannot guarantee the predict precision.

%% file: Tex/7-conclusion.tex
\section{Conclusion}

In this paper, we propose a novel machine learning approach to automatically design superscalar CPUs by learning data dependencies. The predictors are designed using a hardware-friendly model, State-BSD, which consists of a state-selector and a state-speculator to be lightweight and highly precise.  
With State-BSD, we implement the second version of the automated CPU design, i.e., \texttt{QiMeng-CPU-v2}.
To the best of our knowledge, it is the first automated-designed superscalar processor with 380$\times$ performance optimization over state-of-the-art designs, and is comparable to ARM Cortex A53. 
